# Resonant diffusion on solid surfaces

Roumen Tsekov
Department of Physical Chemistry, University of Sofia, 1164 Sofia, Bulgaria

A new approach to Brownian motion of atomic clusters on solid surfaces is developed. The main topic discussed is the dependence of the diffusion coefficient on the fit between the surface static potential and the internal cluster configuration. It is shown this dependence is non-monotonous, which is the essence of the so-called resonant diffusion. Assuming quicker inner motion of the cluster than its translation, adiabatic separation of these variables is possible and a relatively simple expression for the diffusion coefficient is obtained. In this way, the role of cluster vibrations is accounted for, thus leading to a more complex resonance in the cluster surface mobility.

The Brownian motion of atoms and atomic clusters on solid surfaces [1] is of interest for both the pure science and technology. It is relevant to surface coating [2], heterogeneous nucleation [3], catalysis [4], etc. [5]. Nowadays it is realized that the diffusion coefficient of homologies in solids exhibit non-monotone dependence on the molecular size. This phenomenon, called resonant diffusion, has been quantitatively described [6]. The problem here is, if there are resonant effects at the Brownian motion of atomic clusters on solid surfaces. Indeed, similar behavior was experimentally observed for diffusion of Re dimers on tungsten [1], Rh clusters on rhenium [7], etc. Despite the existing theoretical explanations [8, 9], a proper adoption of modern methods for treatment of Brownian motion in modulated structures [10-16] to this important specific problem is required. The basic goal of this paper is to demonstrate a theoretical procedure for description of the resonant diffusion on solid surfaces.

The main tool employed here is a formula derived by Festa and d'Agliano [17] providing possibility to calculate the diffusion coefficient $D$ of a particle moving into a periodic potential

$$1/D = \int_0^1 \beta B(x)\exp[\beta U(x)]d(x/a)\int_0^1 \exp[-\beta U(x)]d(x/a) \qquad (1)$$

where $U$ is the periodic potential with a period $a$ and $\beta = 1/kT$ is the reciprocal thermal energy. Strictly speaking, this formula is valid for the overdamped limit only. However, since we are interested in the large time behavior of the Brownian particle, this condition is always fulfilled. The first problem in the application of Eq. (1) is to determine the friction coefficient $B$. In the frames of the classical statistical mechanics it is possible to obtain a relation between the static potential on the solid surface $U$ and the friction coefficient [13, 15]

$$4\pi\rho c^3 B(x) = (\partial_x^2 U)^2 \qquad (2)$$

where $\rho$ and $c$ are the mass density and sound velocity of the solid, respectively. Expression (2) implies a system, where the anharmonic vibrations do not affect substantially the surface diffusion. Hence, the important part of the lattice dynamics can be accounted well in the harmonic approximation. The Debye spectrum of the phonons is also presumed. The combination of Eqs. (1) and (2) requires additionally a proper expression for the surface potential. A popular model for $U$ is the decoupled cosine potential [18]

$$U = \sum_i A_i \cos(2\pi x_i / a_i) \qquad (3)$$

By variation of the parameters $\{A_i, a_i\}$, one can model both energetic and geometric anisotropy of the surface.

Applying Eq. (3), the potential energy of a $(2n+1)$-chain of a linear rigid homo-polymer moving only in one direction acquires the form

$$U = A[1 + 2\cos(\pi(n+1)l\cos\varphi)\sin(\pi nl\cos\varphi)/\sin(\pi l\cos\varphi)]\cos(2\pi x/a)$$

where $l$ is ratio between the bond length and $a$, $x$ is the mass center coordinate and $\varphi$ is the angle formed by the chain and the lattice x-axis. Introducing this expression into the system of Eqs. (1) and (2) the diffusion coefficient of the chain yields the form

$$D = \rho c^3 a^4 / 2\pi^3 \beta \Delta^2 [I_0(\beta\Delta) + I_2(\beta\Delta)] I_0(\beta\Delta) \qquad (4)$$

where $\Delta \equiv A|1 + 2\cos[\pi(n+1)l\cos\varphi]\sin(\pi nl\cos\varphi)/\sin(\pi l\cos\varphi)|$ and $I_m(\cdot)$ is the m-th modified Bessel function. This expression contains two important limit cases:

i) the Einstein formula $D = \rho c^3 a^4 kT / 2\pi^3 \Delta^2$ at $\beta\Delta \ll 1$;
ii) the Arrhenius formula $D = (\rho c^3 a^4 / 2\pi^2 \Delta) \exp(-2\beta\Delta)$ at $\beta\Delta \gg 1$.

The dependence of the diffusion coefficient on the relative bond length $l$ and the angle $\varphi$ according to Eq. (4) is plotted on Fig. 1. The specific constants of the system are chosen equal to: $\rho = 10^4$ kg/m³, $c = 10^4$ m/s, $A = 10^{-19}$ J, $a = 10^{-10}$ m, $kT = 10^{-20}$ J. As is seen, this dependence is non-monotonous and exhibits a number of minima and maxima. Thus, the resonant diffusion takes place on solid surfaces. It is worth noting that the increase of the number of atoms in the chain leads to more resonances. An important question here is what will occur if the chain is not rigid. Since the general answer is difficult to obtain, we begin with the simplest model, a dimer.

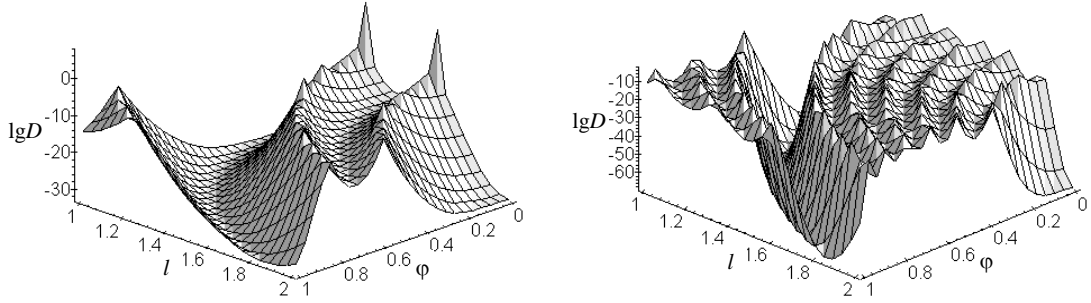

**Fig. 1** Dependence on $l$ and $\varphi$ of diffusion coefficients of trimer and heptamer chains

In a previous paper [18] we examined the role of the rotation on the Brownian motion of rigid dimers. The conclusion was that it affects, in a major way, the resonant character of the diffusion. The present paper aims to elucidate the influence of internal vibrations to the diffusion coefficient of a dimer on a solid surface. The potential energy of a dimer with length $r$ under the field of potential (3) is

$$U = 2A\cos(\pi r/a)\cos(2\pi x/a) \qquad (5)$$

Due to the difference in the characteristic time constants of the dimer translation and vibration, it is possible to apply adiabatic separation of the slow and quick variables. The latter can be considered as freedoms in thermodynamic equilibrium. Hence, the mean effect of the vibrations can be accounted for by the conditional vibration partition function

$$Z = \int_0^\infty \exp[-\beta U(x,r) - \beta V(r)]dr \qquad (6)$$

where $V$ is the internal potential energy of the dimer. A convenient model for $V$ is the harmonic expansion $V = \varepsilon A\pi^2(r/a-l)^2$ where the mean bond length is expressed in lattice parameter units. Using this expression and Eq. (5) one can calculate $Z$ from Eq. (6) and the conditional free energy $F = -kT\ln Z$ of the dimer. Thus calculated $F(x)$ plays the role of an average potential acting on the dimer translation motion and the Festa-d'Agliano formula, extended by Eq. (2), for the present case reads

$$D = 4\pi\rho c^3 / \{\int_0^1 \beta(\partial_x^2 F)^2 \exp[\beta F(x)]d(x/a)\int_0^1 \exp[-\beta F(x)]d(x/a)\} \qquad (7)$$

Because of mathematical difficulties the accomplishment of the above program is frustrated. One can make some reasonable simplifications to better understand the physics. First, if $\varepsilon A > kT$, deviations from the equilibrium length of the oscillator are small and a convenient power series of the external potential energy of the dimer is

$$U = 2A[\cos(\pi l) + \sin(\pi l)\pi(r/a - l) - \cos(\pi l)\pi^2(r/a - l)^2/2 + \cdots]\cos(2\pi x/a)$$

Introducing this expression into Eq. (6) and assuming also $\varepsilon > |\cos(\pi l)|$, yields a simplified formula for the partition function

$$Z = \frac{a}{\pi} \int_{-\pi l}^{\infty} \exp[-2\beta A\cos(\pi l)\cos(2\pi x/a) - 2\beta A\sin(\pi l)\cos(2\pi x/a)q - \varepsilon\beta A q^2]dq$$

To carry out analytically the integration here note the exponent under integration possesses maximum at $q > -\varepsilon^{-1}$. For this reason, the lower bound of the integral can be expanded to infinity without certain loss of correctness. Hence, $F$ is equal to

$$F = 2A\cos(\pi l)\cos(2\pi x/a) - (A/\varepsilon)\sin^2(\pi l)\cos^2(2\pi x/a) \qquad (8)$$

As seen, the *x*-dependence of the free energy is temperature independent, which means that no entropy effects of the vibrations are taken into account.
Unfortunately, performing the integration in Eq. (7) by employing Eq. (8) is still difficult. However, there are two distinguished cases where simple results can be obtained. First, if $l$ differs slightly from a whole number the major part in Eq. (8) is the first one. Hence, the potential is the same as that of a rigid dimer [9] and the activation energy for this case equals to $E_a = 4A|\cos(\pi l)|$. On the contrary, if $l$ is close to a half of a whole number the main term in Eq. (8) is the second part. The two integrals in Eq. (7) can be estimated by the maximal and minimal values of $F$ with respect to $x$. Thus, one can easy conclude that the main contribution to the first integral corresponds to $\cos(2\pi x/a) = 0$, while the second integral can be significantly evaluated at $\cos(2\pi x/a) = 1$. Hence, for the second case the diffusion coefficient of the vibrating dimer can be simply presented as

$$D = [\rho c^3 a^4 \varepsilon^2 / 16\pi^3 \beta A^2 \sin^4(\pi l)]\exp[-\beta A\sin^2(\pi l)/\varepsilon]$$

It is easy to recognize the activation energy of diffusion $E_a = A\sin^2(\pi l)/\varepsilon$.

Finally, assuming additivity of the above contributions, the activation energy of the vibrating dimer in the general case will be given by the expression

$$E_a = A[4|\cos(\pi l)| + \sin^2(\pi l)/\varepsilon] \tag{9}$$

This expression reduces to the former results at their specific values of $l$ and reproduces the known expression for the rigid dimer if $\varepsilon \to \infty$. As seen, the activation energy of the vibrating dimer is always larger than that of the corresponding rigid dimer. The reason for this unexpected result could be elucidated using Eq. (8). The physics of this phenomenon is that during the dimer vibrations, it passes many times through states with different fit to the surface potential. But in the quadratic contribution to the free energy the resulting resonances cumulate and thus always lead to an increase in the activation energy. The dependence of $D$ from $\varepsilon$ and the average bond length is plotted in Fig. 2. It is strongly resonant as a function of the mean dimer length and substantially modulated by the dimer vibrations.

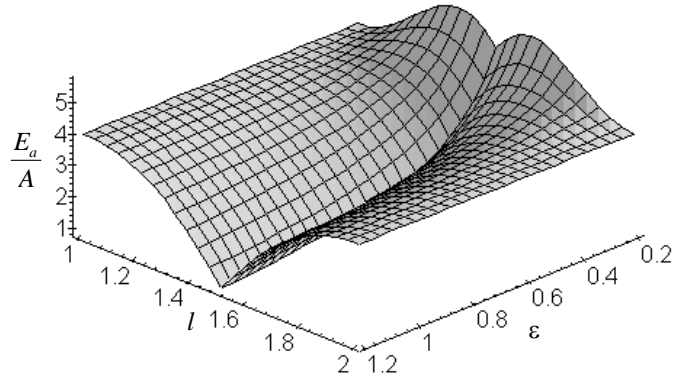

**Fig. 2** Dependence of the activation energy of a vibrating dimer on $l$ and $\varepsilon$

In the present paper an adaptation of the Festa-d'Agliano formula to the diffusion of clusters on solid surfaces is presented. A formula for calculation of the diffusion coefficients of linear rigid polymers is derived which clearly expresses non-monotonous dependence from the geometric parameters of the chain. An important feature is the diffusion coefficient strongly depends upon the number of atoms in the chain, which can be utilized for successful separation of mixtures from homologies. The role of internal vibrations of adsorbed dimers on their diffusion is also studied in this paper. In the case of a strong chemical interaction between the atoms of the dimer, the present theory rederives our previous results for rigid dimers. In the opposite case the internal vibrations strongly affect the diffusion and lead to an increase in the activation energy of the dimer by means of accumulation of the effects of any moment state of the system. This is an unexpected phenomenon, since the usual notion is that the vibrations could lead to a better energetic fitness between the dimer and solid surface at any moment. The developed concept is open for any other application. A future goal is to calculate simultaneously the contribution of rotation and vibration of a dimer to its diffusion coefficient. Also it is more practical to apply our procedure to the Brownian motion of a small cluster with arbitrary shape, a problem important to technology. Additional complications could arise from important contribution of inertial [19] and quantum [16, 20] effects.

1. G. Erhlich and K. Stolt, *Annu. Rev. Phys. Chem.* **31**, 603 (1980)
2. G. Ehrlich, *Surf. Sci.* **331-333**, 865 (1995)
3. G. Kellog, *Surf. Sci. Rep.* **21**, 1 (1994)
4. G. Vayssilov, *Adv. Colloid Interface Sci.* **43**, 51 (1993)
5. K. Christmann, *Introduction to Surface Physical Chemistry* (Springer, New York, 1991)
6. R. Tsekov and E. Ruckenstein, *J. Chem. Phys.* **100**, 3808 (1994)
7. G. Kellog, *Phys. Rev. Lett.* **73**, 1833 (1994)
8. R. Wang and K. Fichthorm, *Phys. Rev. B* **48**, 18288 (1993)
9. E. Ruckenstein and R. Tsekov, *J. Chem. Phys*. **100**, 7696 (1994)
10. W. Kleppmann and R. Zeyher, *Phys. Rev. B* **22**, 6044 (1980)
11. K. Golden, S. Goldstein and J. Lebowitz, *Phys. Rev. Lett.* **55**, 2629 (1985)
12. R. Ferrando, R. Spadacini, G. Tommei and G. Caratti, *Physica A* **195**, 506 (1993)
13. R. Tsekov and E. Ruckenstein, *J. Chem. Phys.* **100**, 1450 (1994)
14. Y. Georgievskii and E. Pollak, *Phys. Rev. E* **49**, 5098 (1994)
15. R. Tsekov and E. Ruckenstein, *J. Chem. Phys.* **101**, 7844 (1994)
16. G. Vayssilov, *Adv. Colloid Interface Sci.* **57**, 123 (1995)
17. R. Festa and E. d'Agliano, *Physica A* **90**, 229 (1978)
18. R. Tsekov and E. Ruckenstein, *Surf. Sci*. **344**, 175 (1995)
19. R. Tsekov, *Ann. Univ. Sofia, Fac. Chem.* **88** (1), 57 (1995)
20. R. Tsekov, *J. Phys. A: Math. Gen.* **28**, L557 (1995)